\documentstyle[12pt, fleqn]{article}
\begin{document}
\parskip=12pt
\vspace{5cm}
{\begin{center}
{\Large{SU(2) $\times$ U(1) Gauge  Gravity }}
\end{center}
\vspace{5cm}
\begin{center}
{{\sc H. Dehnen and E. Hitzer} \\
{\it Physics Department\\
University of Konstanz \\
Box 5560\\
D-78434 Konstanz}}
\end{center}
\newpage

\section*{Abstract}

We propose a Lorentz-covariant Yang-Mills ``spin-gauge'' theory, where the
function valued Pauli matrices play the role of a non-scalar Higgs-field. As
symmetry group we choose $SU(2) \times U(1)$ of the 2-spinors describing
particle/antiparticle states. After symmetry breaking a non-scalar
Lorentz-covariant  Higgs-field gravity appears, which can be interpreted
within a classical limit as Einstein's metrical theory of gravity, where we
restrict ourselves in a first step to its linearized version. \newpage

\section*{I. Introduction}
In a previous paper (Dehnen, Hitzer, 1994) we proposed a unitary gauge theory
of gravity in view of the possibility of quantization of gravity and its
unification with the other physical interactions. In this theory, where the
subgroup $SU(2) \times U(1)$  of the
 unitary transformations of Dirac's $\gamma $-matrices  between their
different representations [internal spin group (see also Drechsler, 1988 and
Bade, Jehle, 1953;
 cf. also Laporte and Uhlenbeck, 1931; Barut and Mc Ewan, 1984)] is gauged,
 the $\gamma $-matrices become function valued. Because the $\gamma$-matrices
can be understood as the square root of the metric our gauge group is the
unitary group belonging to the square root of the metric. Taking the function
valued $\gamma $-matrices as  true field
 variables with a Higgs-Lagrange density, and this because also
 the $\gamma $-matrices possess a non-trivial ground-state, namely
 the usual constant standard representations, we got a  unitary spin-gauge
theory with
 Dirac's $\gamma $-matrices as Higgs-fields.  After spontaneous symmetry
breaking a non-scalar Higgs-gravity
 appears, which
 can be connected in a classical limit with Einstein's gravity,
 where we restricted ourselves in the first step for reasons of simplicity to
the linear theory.

 The essential points are the following: The theory is from the
 beginning only Lorentz-covariant. After symmetry breaking
 the action of the excited
$\gamma $-Higgs-field on the fermions in the Minkowski
space-time can be reinterpreted as if there would exist non-euclidean
space-time connections and a non-euclidean metric
 (effective metric), in which the fermions move freely; then the
 deviation from the Minkowski space-time describes classical
 gravity. Simultaneously the  gravitational constant is produced
 only by the symmetry
 breaking   whereby
 the gauge-bosons get masses of the order of the Planck-mass and
 can therefore be neglected in the low energy limit; but in the
 high energy limit ($ \simeq 10 ^{19} $ GeV) an additional
 "strong" gravitational
 interaction exists.

However we found also a richer space-time
 geometrical structure  than only a Riemannian one. We got beside an
effective metric
 also an effective  non-metricity, whereas an effective torsion did not
appear.

The aim of the present paper is to avoid non-metricity and
 torsion which is indeed possible by changing the Lagrange-density. We
develop at first a quantum mechanical description of the gravitational
interaction between fermionic elementary particles and  arrive subsequently
in the linearised
 version  of the classical limit exactly at Einstein's linearized theory.
However, instead of starting from
 Dirac's 4-spinor formalism it is more appropriate to
 begin with the 2-spinor description of massless spin-1/2 fermions, because
 then the gauge-group $SU(2) \times  U(1)$ is irreducible. In
 consequence of this several considerations become much clearer than in the
 foregoing paper especially the symmetry breaking and the
 transition to the classical macroscopic limit. Moreover there is a further
essential reason for starting with the 2-spinors. There exist strong hints
that also for antiparticles the weak
 equivalence principle
$\overline{m}_g \equiv \overline{m}_i (\equiv m_i ) $ is valid (Nieto,
Goldman, 1991; see also Morrison,
 1958 and  Ebner, Dehnen, 1993). Then particle and antiparticle
 are indistinguishable with respect to gravity (and only with respect to
gravity) as it is also postulated classically by general
 relativity. Therefore it is consequent on the quantum mechanical
 level to combine particle and antiparticle as a particle-doublet, on which
the $SU(2) \times U(1)$  group acts. This procedure,
 which is exactly the same as that of electroweak and strong
 interaction or their unifications, is possible by choosing the
 2-spinors following from the chiral decomposition of Dirac's theory. For
this reasons we think that the gauge group in question
 is indeed that of microscopic gravity, from which Einstein's
 macroscopic gravity follows in the classical limit as an effective field.

\section*{II. The Basic Concept}

Starting from the Lagrange-density of massless spin-1/2 particles within the
4-spinor formalism $(\hbar = 1, c = 1)$
$$
{\cal{L}} _M = \frac{i}{2} \overline{\psi } \gamma ^\mu
\partial _ \mu  \psi  + h. c
\eqno (2.1)
$$
with the anticommutator relation
$$
\gamma ^{( \mu } \gamma ^{\nu )} = \eta ^{\mu \nu } {\bf 1}
\eqno (2.1a)
$$
$( \eta ^{\mu \nu } = \eta _ {\mu \nu } = \mbox{diag} (+1, -1, -1, -1)$
Minkowski metric) we go over to a 2-spinor
description using the chiral decomposition:
$$
\psi = \left(
\begin{array}{l}
\chi _R \\
 \varphi _L
\end{array}
\right)  , \quad
\gamma ^\mu  =
\left(
\begin{array}{ll}
0 & \sigma ^\mu _ L \\
\sigma ^\mu _ R & 0
\end{array}
\right)
\eqno (2.2)
$$
where
$$
\sigma ^\mu _L = ( {\bf 1} , - \sigma ^i ) , \quad \sigma ^\mu _R =
({\bf 1} , \sigma ^i)
\eqno (2.2a)
$$
($\sigma^i , i = 1,2,3 $ Pauli-matrices). Herein the 2-spinor
 $\chi _R$ represents the right-handed particle- and left-handed
antiparticle-states and $\varphi _L$ the left-handed particle- and
right-handed antiparticle-states.  Inserting
 (2.2) and (2.2a) into (2.1) and (2.1a) we find the matter Lagrange-density
$$
{\cal{L}}_M = \frac{i}{2} (\chi _R ^{\dagger} \sigma ^\mu _R \partial _ \mu
\chi _R +
\varphi _L ^{\dagger}  \sigma ^\mu _L \partial _\mu \varphi _L ) + h.c.
\eqno (2.3)
$$
and the "anticommutator" relations:
$$
\sigma ^{(\mu }_L \sigma ^{\nu )}_R = \eta ^{\mu \nu } {\bf 1},
\quad \sigma ^{(\mu }_R \sigma ^{\nu )} _L = \eta ^{\mu \nu } {\bf 1} .
\eqno (2.4)
$$

The Lagrange-density (2.3) and the relations (2.4) are
invariant or covariant under the global $SU(2) \times U(1)$ transformations
$$
\begin{array}{lll}
\chi _R ^\prime & = &  U \chi _R , \quad \varphi _L ^\prime = U \varphi _L \\
[1ex]
\sigma ^{\mu \prime} _R & = &  U \sigma ^\mu _R U^{-1} , \quad \sigma ^{\mu
\prime}_L = U \sigma ^\mu _L U^{-1}
\end{array}
\eqno (2.5)
$$
with
$$
\begin{array}{lll}
U & = &  e^{i \lambda _a \tau ^a} , \quad \tau ^a = \frac{1}{2}\sigma ^a ,
\quad \sigma ^a  =  ({\bf 1} , \sigma ^i ) \\
(a & = & 0,1,2,3; \quad i = 1,2,3 )
\end{array}
\eqno (2.5a)
$$
($\sigma ^i$ Pauli-matrices as generators of the
transformation group $SU(2)$), where $\lambda _a = $ const. (real valued) and
the
generators $\tau ^a$ fulfil the commutator relation
$$
[ \tau ^a , \tau ^b ] = i \epsilon ^{ab} {}_c \tau  ^c
\eqno (2.5b)
$$
($ \epsilon ^{ab} {}_c $ is the Levi-Civita symbol with the
additional property to be zero, if $a, b$ or $c$ are zero).

Now we gauge this group by setting $\lambda _a = \lambda _ a (x ^\mu )$ (real
valued functions). Then the two states of $\chi _R$ and of $\varphi _L$, i.e.
the (right- and left-handed) particles-antiparticles, which are mixed by this
gauge group according to (2.5), are indistinguishable with respect to the
resulting interaction (c.f. introduction), whereby the lepton- and
baryon-number conservation is violated because of the possibility of
particle-antiparticle transitions (see below). The invariance of the
 Lagrange-density (2.3) is guaranteed in future by
substituting the usual partial derivative through the
covariant one: {\footnote{$| \mu $ denotes the partial
derivative with respect to the coordinate $x ^\mu $.}}
$$
\begin{array}{lll}
D_ \mu & = &  \partial _\mu  + i g \omega _\mu \\
\omega _ \mu ^{\prime} & = & U \omega _ \mu U^{-1} + \frac{i}{g} U_ {|\mu }
U^{-1}
\end{array}
\eqno (2.6)
$$
($g $  dimensionless gauge-coupling constant). The real
valued gauge potentials $ \omega _ {\mu a} $ are defined by
$$
\omega _ \mu = \omega _ {\mu a} \tau ^a .
\eqno (2.6a)
$$
However simultaneously the $\sigma ^\mu _ R $- and
$\sigma ^\mu _L $-matrices become function valued because
of the transformation law (2.5). We denote these function
valued matrices $\, \tilde \sigma ^\mu _R $ and $\, \tilde \sigma ^\mu _L$
respectively (bear this notation in mind!);
they are Hermitean and therefore in the adjoint representation.

Furthermore the Lagrange-density (2.3) must be supplemented by
two parts: First by a part for the gauge-potentials
$\omega _ {\mu a}$ and secondly by a part for the $ \, \tilde \sigma ^\mu _
R$- ,  $\, \tilde \sigma ^\mu  _ L$-matrix functions.
We choose for the latter ones a Higgs-field Lagrange density, because $\,
\tilde \sigma ^\mu _R $ and
$\, \tilde \sigma ^\mu _L$ possess a natural non-trivial ground-state
given by (2.2a). Relations  (2.4) then only apply to the ground-states. In
general the
function valued matrices $\, \tilde \sigma ^\mu _R $ and
$\, \tilde \sigma ^\mu _L$  later  on will define  an \underline{effective}
function valued (non Euclidean)
metric $g^{\mu \nu }$.

Thus the total Lagrange-density consists of three
minimally coupled Lorentz-  and gauge-invariant
real valued parts:
$$
{\cal{L}}= {\cal{L}}_M (\psi ) +
{\cal{L}}_F (\omega ) + {\cal{L}}_H (\, \tilde \sigma ).
\eqno (2.7)
$$
Beginning with the third brand new part, ${\cal{L}}_H (\, \tilde \sigma )$
belongs to the $\, \tilde \sigma ^\mu _ {R, L}$-Higgs-fields and  we choose
for this:
$$
{\cal{L}}_H (\, \tilde \sigma ) = \mbox{tr}
(D_ \alpha \, \tilde \sigma _ {\mu R} ) (D^\alpha \, \tilde \sigma ^\mu _L )
- \mbox{tr} (D_ \alpha \, \tilde \sigma _ {\mu R} )
(D^\mu \, \tilde \sigma ^\alpha _L ) -
$$
$$
- \mbox{tr} (D_ \alpha \, \tilde \sigma ^\alpha _R)
(D_ \beta
\, \tilde \sigma ^\beta _ L) - V( \, \tilde \sigma )
\eqno (2.8)
$$
$$
- k [ \varphi _L ^{\dagger} \, \tilde \sigma ^\mu _L \, \tilde \sigma  _{\mu
R} \chi _R +
\chi _R ^{\dagger} \, \tilde \sigma ^\mu _R    \, \tilde \sigma _ {\mu L}
\varphi _L ] ,
$$
where the last term is a Yukawa-coupling term ($k$ is a
dimensionless coupling constant) for generating the
mass of the fermions by the $\, \tilde \sigma _ {R,L}^\mu $-Higgs-fields
after spontaneous symmetry breaking.{\footnote{It is worth to note, that this
Yukawa coupling term is necessary for arriving at Einstein's theory in the
classical limit.}} The
Higgs-potential $V(\, \tilde \sigma )$ takes the form:
$$
V(\, \tilde \sigma ) = \mu ^2 \mbox{tr}(  \, \tilde \sigma ^\mu _L \, \tilde
\sigma _ {\mu R})  + \frac{\lambda }{12} (\mbox{tr}  \, \tilde \sigma ^\mu _L
  \, \tilde \sigma _{\mu R} )^2.
\eqno (2.8a)
$$
($\lambda > 0$ and $\mu ^2 < 0 $ are real valued constants;
$\lambda $ is dimensionless, $\mu ^2$ has the dimension
of a mass square). In the kinetic part of (2.8) all
 possible different combinations between the
derivatives of $\, \tilde \sigma ^\mu _R$ and $\,
\tilde \sigma _L ^\mu $ are taken into account.

The second term on the right hand side of (2.7) is that of the gauge
potentials $\omega _ {\mu a}$ and has the usual form:
$$
{\cal{L}}_F (\omega ) = - \frac{1}{16 \pi } F_ {\mu \nu a}
F^{\mu \nu } {}_ b s^{ab} ,
$$
where $s ^{ab} $ is the group-metric of SU(2) $\times$
U(1) and can be taken as $\delta ^{ab}$ . The gauge-field
 strengths are defined in the usual manner by
$$
{\cal{F}} _ {\mu \nu } = \frac{1}{ig} [D_ \mu , D_ \nu ] =
F_ {\mu \nu a } \tau ^a
\eqno (2.9a)
$$
with
$$
F_ {\mu \nu a} = \omega _ {\nu a | \mu } -
\omega _ {\mu a | \nu } -
g \epsilon _a {}^{bc} \omega _ {\mu b }\omega _ {\nu c} .
\eqno (2.9b)
$$

The first term in the Lagrange density in (2.7) concerns the fermionic matter

fields and is given by the gauge invariant modification of (2.3):
$$
{\cal{L}}_M (\psi ) =
\frac{i}{2} \left\{
\chi ^{\dagger} _R \, \tilde \sigma ^\mu _ R D_ \mu \chi _R -
( D_ \mu \chi _R ) ^{\dagger} \, \tilde \sigma ^\mu _R \chi _R  \right.
$$
$$
+ \varphi _L ^{\dagger} \, \tilde \sigma ^\mu _L D_ \mu \varphi _L -
\left. (D_ \mu \varphi _L ) ^{\dagger} \, \tilde \sigma ^\mu _L \varphi _L
\right\} .
\eqno (2.10)
$$
We note that the total Lagrangian (2.7) contains no dimensional
 parameters except  $\mu ^2 $ in the
 Higgs-potential (2.8a), which has the dimension of
a mass square. Because in the following by the symmetry breaking only one
sort of mass is generated, the weak equivalence principle for particles and
antiparticles is guaranted from the very beginning in the general form $m_i
\equiv m_g \equiv \overline{m}_i \equiv \overline{m}_g \equiv m$, see (3.5).

The field equations following from the action principle
associated with (2.7) are the generalized 2-spinor equations
$$
i \, \tilde \sigma ^\mu _R D_ \mu  \chi _R +
\frac{i}{2} (D_ \mu \, \tilde \sigma ^\mu _R ) \chi _R -
k \, \tilde \sigma ^\mu _ R \, \tilde \sigma _ {\mu L} \varphi _L = 0,
\eqno (2.11a)
$$
$$
i \, \tilde \sigma ^\mu _L D_ \mu \varphi _L +
\frac{i}{2} (D_ \mu \, \tilde \sigma ^\mu _L ) \varphi _L
 - k \, \tilde \sigma ^\mu  _ L \, \tilde \sigma _ {\mu R } \chi _R = 0
\eqno (2.11b)
$$
as well as their adjoint equations, and the inhomogeneous Yang-Mills
equations:
$$
\partial _\nu F^{\nu \mu } _a +
g \epsilon _a {}^{bc} F_ b ^{\nu \mu } \omega _ {\nu c} =
4 \pi j ^\mu _a
\eqno (2.12)
$$
with the gauge-currents
$$
j^\mu _a = j^\mu _ a (\psi ) + j^\mu _a (\, \tilde \sigma )
\eqno (2.12a)
$$
consisting of a real-valued matter part:
$$
j^\mu _a (\psi ) =
\frac{g}{2}
\left[ \chi ^{\dagger} _R \left\{  \, \tilde \sigma ^\mu _R , \tau _a
\right\} \chi _R + \varphi ^{\dagger} _L \left\{  \, \tilde \sigma ^\mu _L ,
\tau _a \right\} \varphi _L \right]
\eqno (2.12b)
$$
and a real valued $\, \tilde \sigma $-Higgs-field part:
$$
j^\mu _a (\, \tilde \sigma ) =
ig \mbox{tr}  \left\{  [\, \tilde \sigma _ {\alpha R} , \tau _a ]
D^\mu \, \tilde \sigma ^\alpha _ L  +
\right.
  [\, \tilde \sigma _ { \alpha L} , \tau _a ] D^\mu \, \tilde \sigma ^\alpha
_R  -
$$
$$
- [\, \tilde \sigma _ {\alpha R} , \tau _a ] D^\alpha
\, \tilde \sigma ^\mu _L
-   [\, \tilde \sigma _ {\alpha L} , \tau _a ] D^\alpha
\, \tilde \sigma ^\mu _R  -
\eqno (2.12c)
$$
$$
-  [\, \tilde \sigma ^\mu  _R , \tau _a ] D_ \alpha \, \tilde \sigma ^\alpha
_L
\left.
-  [\, \tilde \sigma ^\mu _L , \tau _a] D_ \alpha \, \tilde \sigma ^\alpha _R
 \right\} .
$$
Finally we get the Higgs-field
equations for $\, \tilde \sigma ^\mu _R $ and
$\, \tilde \sigma ^\mu _L $ respectively:
$$
(D_ \alpha D^\alpha \, \tilde \sigma ^\mu _R )  _A {}^B  -
(D_ \alpha D^\mu \, \tilde \sigma ^\alpha _R )   _A {}^B
- ( D^\mu  D_ \alpha \, \tilde \sigma ^\alpha _R ) _ A {}^B +
$$
$$ +
[\mu ^2 + \frac{\lambda }{6} \mbox{tr}
(\, \tilde \sigma ^\alpha _ L \, \tilde \sigma _ {\alpha R})]
\, \tilde \sigma ^\mu _ {RA} {}^B =
$$
$$
= \frac{i}{2} [\varphi _L ^{{\dagger} B } (D^\mu \varphi _L )_A
- (D^\mu  \varphi _L ) ^{{\dagger} B} \varphi _ {LA} ] -
\eqno (2.13)
$$
$$
- k [ \varphi _L ^{{\dagger} B} (\, \tilde \sigma  ^\mu _R \chi _R )_ A   +
(\chi _R ^{\dagger} \, \tilde \sigma ^\mu  _R )^B \varphi _ {LA}]
$$
and
$$
(D_ \alpha D^\alpha \, \tilde \sigma ^\mu _L ) _A {}^B -
(D_ \alpha D^\mu  \, \tilde \sigma ^\alpha _L ) _ A {}^B  - (D^\mu D_ \alpha
\, \tilde \sigma ^\alpha _L ) _A {}^B +
$$
$$
+
[\mu ^2 + \frac{\lambda }{6} \mbox{tr} \, \tilde \sigma ^\alpha _L \, \tilde
\sigma _ {\alpha R}] \sigma ^\mu _ {L A} {}^B =
$$
$$
= \frac{i}{2} [\chi _R ^{{\dagger} B} (D^\mu \chi _R ) _A -
(D^\mu  \chi _R )^{{\dagger} B} \chi _ {RA} ] -
\eqno (2.14)
$$
$$
- k [\chi ^{{\dagger} B} _R (\, \tilde \sigma ^\mu _L \varphi _L )_ A ] +
(\varphi _L ^{\dagger} \, \tilde \sigma ^\mu _L )^B \chi _ {RA} ].
$$
Herein the lower capital latin index $A$ and the upper index $B$ denote the
contragradiently transformed rows and columns  of the 2-spinorial quantities
respectively. The homogeneous Yang-Mills equation following from the
Jacobi-identity reads:
$$
\partial _ {[ \mu } F_ {\nu \lambda ] a} +
g \epsilon _a {}^{bc} \omega _ {b [ \mu } F_ {\nu \lambda ] c} = 0.
\eqno (2.15)
$$
The right hand sides of (2.13) and (2.14) are  Hermitean and therefore $\,
\tilde \sigma ^\mu  _R$ and $\, \tilde \sigma ^\mu _L $ remain  also in
consequence of the field equations Hermitean.

Finally we note the conservation laws in consequence of the
invariance-structure of the Lagrangian, valid modulo the field equations.
First, from (2.12) the gauge-current conservation follows immediately:
$$
\partial _\mu (j ^\mu _a + \frac{g}{4 \pi } \epsilon _a {}^{bc} F^{\mu \nu }
{}_b
\omega _{\nu c} ) = 0.
\eqno (2.16)
$$
Secondly, the energy-momentum law takes the form
$$
\partial _\nu T_ \mu {}^\nu = 0,
\eqno (2.17)
$$
where $T _ \mu {}^\nu$ is the gauge-invariant canonical energy-momentum
tensor consisting of three real valued parts corresponding to (2.7):
$$
T_ \mu {}^\nu  = T_ \mu {}^\nu (\psi ) +
T_ \mu {}^\nu  (\omega ) + T  _\mu {}^\nu (\, \tilde \sigma ).
\eqno (2.18)
$$
Herein $T_ \mu {}^\nu (\psi )$ has a right-  and a left-handed part:
$$
T_ \mu {}^\nu  (\psi ) =
T_ \mu {}^\nu (\chi _R) +
T_ \mu {}^\nu (\varphi _L)
\eqno (2.19)
$$
with (modulo the field equations (2.11))
$$
T_ \mu {}^\nu (\chi _R) =
\frac{i}{2} [\chi _R ^{\dagger} \, \tilde \sigma ^\nu _R D_ \mu \chi _R -
(D_ \mu \chi _R )^{\dagger} \, \tilde \sigma ^\nu _R \chi _R ]
\eqno (2.19a)
$$
and
$$
T_ \mu {}^\nu (\varphi _L ) = \frac{i}{2} [ \varphi _L ^{\dagger}
\, \tilde \sigma _L ^\nu D_ \mu \varphi _ L -
(D_ \mu \varphi _L ) ^{\dagger} \, \tilde \sigma ^\nu _L \varphi _L ] .
\eqno (2.19b)
$$
The second term on the right hand side of (2.18) is given as usual by
$$
T_ \mu {}^\nu  (\omega  ) =
- \frac{1}{4 \pi } [F ^a _ {\mu \alpha }  F^{\nu \alpha } _ a
- \frac{1}{4} \delta ^\nu _ \mu F^a _ {\alpha \beta } F^{\alpha \beta } _a ]
\eqno (2.20)
$$
and the last term possesses the form:
$$
T _ \mu {}^\nu (\, \tilde \sigma ) =
\mbox{tr} [(D^\nu  \, \tilde \sigma ^\alpha _L )( D_ \mu \, \tilde \sigma _
{\alpha R}) -
(D^\alpha \, \tilde \sigma ^\nu _L) (D_ \mu \, \tilde \sigma _ {\alpha R}) -
$$
$$
- (D_ \alpha \, \tilde \sigma ^\alpha _L ) (D_ \mu \, \tilde \sigma ^\nu _R )
+ (D^\nu \, \tilde \sigma ^\alpha _R)
(D_ \mu \, \tilde \sigma _ {\alpha L} ) -
$$
$$
- (D^\alpha \, \tilde \sigma ^\nu _ R) (D_ \mu \, \tilde \sigma _ {\alpha L})
-
(D_ \alpha \, \tilde \sigma ^\alpha _R ) (D_ \mu \, \tilde \sigma ^\nu _L)] -

\eqno (2.21)
$$
$$
- \delta ^\nu _\mu  \Bigl\{  \mbox{tr} [(D_ \alpha \, \tilde \sigma _ {\beta
R})
(D^\alpha \, \tilde \sigma ^\beta _L) -
(D_ \alpha \tilde \sigma _{\beta R}) (D^\beta  \tilde \sigma ^\alpha _L) -
(D_ \alpha  \tilde \sigma ^\alpha _R) (D_ \beta \tilde \sigma ^\beta _L ) ] -

$$
$$
- [ \mu ^2 \mbox{tr} (\, \tilde \sigma ^\alpha _L \, \tilde \sigma _ {\alpha
R})
+  \frac{\lambda }{12} (\mbox{tr}  \tilde \sigma ^\alpha _L \, \tilde \sigma
_ {\alpha R}) ^2 ] \Bigr\} .
$$

By insertion of (2.19), (2.20) and (2.21) into (2.18) one obtains from (2.17)
the momentum-law for fermions. After substituting the covariant D'Alembertian
of $\, \tilde \sigma ^\mu _R$ and $\, \tilde \sigma ^\mu _L$ by the field
equations (2.13) and (2.14) one finds using the Yang-Mills-equations (2.12)
and (2.15):
$$
\partial _\nu T _\mu {}^\nu (\psi ) =
- \frac{i}{2} \left[ \varphi _L ^{\dagger}
(D_ \mu \, \tilde \sigma ^\alpha _L) D_ \alpha \varphi _L  \right. - (D_
\alpha \varphi _L) ^{\dagger} (D_ \mu \, \tilde \sigma ^\alpha _L ) \varphi
_L +
$$
$$
 +
\left.  \chi ^{\dagger} _R (D_ \mu \tilde \sigma ^\alpha _R )
D_ \alpha \chi _R -
 (D_ \alpha \chi _R) ^{\dagger} (D_ \mu \tilde \sigma ^\alpha _R )
 \chi _R \right] +
\eqno (2.22)
$$
$$
+ k \left\{   \varphi _L ^{\dagger} \left[ D_ \mu ( {\, \tilde \sigma}
^\alpha _L  \tilde \sigma _ {\alpha R} ) \right] \chi _R +
\chi _R ^{\dagger} \left[ D_ \mu ( {\, \tilde \sigma} ^\alpha _R  {\, \tilde
\sigma }_ {\alpha L} ) \right]  \varphi _L \right\}
$$
$$
+ F_ {\mu \nu a} j^{\mu a} (\psi ) .
$$
On the right hand side one recognizes the Lorentz-like forces of the
gauge-fields and the forces of the $\, \tilde \sigma ^\mu _R$- and $\, \tilde
\sigma ^\mu _L $-Higgs-fields.
Although the field equations seem to be very complicated, there exists a very
transparent structure and physical meaning.

First of all on the right hand sides of (2.13) and (2.14) there appear the
fermionic energy momentum tensors (2.19a) and (2.19b), i.e. $T_ \mu {}^\nu
(\chi _R)$  and $T_ \mu {}^\nu  (\varphi _L) $ in their spinor valued form as
sources of the $\, \tilde \sigma ^\mu _L$- and $\, \tilde \sigma ^\mu
_R$-fields respectively and these fields act back by their gradients on the
fermions in their field equations (2.11) and the momentum law (2.22). The
structure is exactly that of an attractive gravitational interaction with the
energy momentum tensor as source. However we have two different gravitational
fields, namely $\, \tilde \sigma ^\mu _R $ and $\, \tilde \sigma ^\mu _L$.
Only in the classical limit, where no distinction of right- or left-handed
states exists, we get a universal gravitational interaction which can be
described finally by a universal effective non Euclidean metric.

\section*{III. Spontaneous Symmetry Breaking and the Fermionic and Bosonic
Masses}

Although one can recognize the gravitational structure already
in the foregoing chapter, the physical interpretation will
be much clearer after symmetry breaking. The minimum of the
energy-momentum tensor (2.18) in absence of matter- and
gauge-fields coincides with the minimum of  the Higgs-potential (2.8a)
 defined by:
$$
\mbox{tr} (\stackrel{(0) }{\, \tilde \sigma ^\mu} _L
\stackrel{(0)}{\, \tilde \sigma }_ {\mu R} ) =
- \frac{6 \mu ^2}{\lambda } =
\frac{1}{2} v^2 \quad (\mu ^2 < 0) .
\eqno (3.1)
$$
Simultaneously, herewith all field-equations (2.11) up to
(2.15) are fulfilled. The ground-states $\stackrel{(0) }{\, \tilde \sigma^\mu
 }  _L$ and $\stackrel{(0)}{\, \tilde \sigma }_ {\mu R} $ must be
proportional to (2.2a). In view
of (2.4) one then finds from (3.1):
$$
\stackrel{(0)}{\, \tilde \sigma^\mu }_L =
\frac{v}{4} \sigma ^\mu _L ,
\quad \stackrel{(0)}{\, \tilde \sigma ^\mu} _R  = \frac{v}{4}
 \sigma ^\mu _R .
\eqno (3.2)
$$
Because $\, \tilde \sigma ^\mu _L$ and $\, \tilde \sigma ^\mu _R$ are
Hermitean (adjoint representation) we can reduce them
to the ground-states as follows:
$$
\, \tilde \sigma ^\mu _L = h^\mu {} _ {\nu L}
\stackrel{(0)}{\, \tilde \sigma ^\nu} _L , \quad
\, \tilde \sigma ^\mu _R = h^\mu {}_ {\nu R}
\stackrel{(0)}{\, \tilde \sigma^\nu } _R
\eqno (3.3)
$$
with the real valued fields $h^\mu {}_ {\nu L}$ and
$h^\mu {}_ {\nu R}$ ; they have the structure:
$$
h^\mu {}_ {\nu L} = \delta ^\mu {}_ \nu +
\epsilon ^\mu {}_ {\nu L} ,
\quad
h^\mu {}_ {\nu R} = \delta ^\mu {}_\nu +
\epsilon ^\mu {}_ {\nu R}
\eqno (3.3a)
$$
with the excited Higgs-fields $\epsilon ^\mu {}_ {\nu L}$ and $\epsilon ^\mu
{}_ {\nu R}$. In addition we set
$$
\chi _R = \frac{2}{\sqrt v} \chi _ {RD} ,
\quad \varphi _L =
\frac{2}{\sqrt v} \varphi _ {LD} ,
\eqno (3.4)
$$
so that the real Dirac-spinor $\psi _D$ reads now
(see (2.2)):
$$
\psi _D ={ \chi _{RD} \choose \varphi _{LD}} .
\eqno (3.4a)
$$
In this way the dimension of $\, \tilde \sigma ^\mu _L $, $ \, \tilde \sigma
^\mu _R$ is compensated.

Insertion of (3.2) and (3.4) into the Yukawa-coupling
term of the Lagrange-density (2.8) using (2.4) leads immediately
to the single fermionic mass $m$ :
$$
m = kv .
\eqno (3.5)
$$
On the other hand the mass of the gauge-bosons $\omega _ {\mu a}$ results
from the Higgs-current (2.12c) in its lowest order; insertion of (3.2) gives:

$$
- j^\mu {}_a (\stackrel{(0)}{\, \tilde \sigma }) =
[ \frac{1}{2}\eta ^{\mu \nu } M^2  {}_\alpha {}^\alpha {}_ {ab} -
(M^{2 \mu \nu } {}_ {ab} + M^{2 \nu \mu } {}_ {ab} -
\frac{1}{2} \eta ^{\mu \nu }  M^2 {}_ \alpha {}^\alpha {}_ {ab} ) ] \omega
_\nu {}^{b}
\eqno (3.6)
$$
with the mass-square matrix
$$
M^{2 \mu  \nu } {}_ { ab} =
- \frac{g^2 v^2}{16} \left\{  \mbox{tr}
([ \sigma ^\mu _R , \tau _a] [\sigma ^\nu _L , \tau _b] )+
\mbox{tr}( [\sigma ^\mu _L , \tau _a] [\sigma ^\nu _R , \tau _b]) \right\} .
\eqno (3.6a)
$$
The bracket in (3.6) consists of two parts symmetric
with respect to $\mu , \nu $ as well as to $a, b$. The first
term represents the trace of $M^{2 \mu \nu } {}_ {ab}$ referring to $\mu ,
\nu $
and the second one is the traceless symmetric part of it
 giving rise to an anisotropy of the
(effective) mass of the gauge bosons. If we later on identify
the gravitational interaction mentioned in the foregoing
chapter in its classical limit with Einstein's macroscopic metrical gravity,
we will find (see (5.34))  $v^2 =
(2 \pi G)^{-1} $ (G Newton's gravitational constant).
Therefore the $SU(2)$-gauge-bosons ($ a = i = 1,2,3)$ get
masses of the order of the Planck-mass $M_ {Pl} =
 1/ \sqrt{G}$  $ (\stackrel{\wedge}{=}   10^{19}$ GeV )
according to (3.6a) and can be neglected in the low
energy limit. In the higher energy ranges however an
 additional "strong gravitational" interaction exists
mediated by the three very massive $\omega _ {\mu i}$- bosons, which violates
in view of the transition-currents (2.12b) the lepton- and baryon-number
conservation.  With respect to the $SU(2)$-group only global transformations
are allowed from now. On
the other hand the U(1)-gauge boson ($a = 0)$ remains
massless in view of (3.6a) and (2.5a). Therefore the
U(1)-gauge group represents the rest-symmetry group and can be identified
with that of the
weak hypercharge, so that here a unification with the
electroweak interaction intrudes by setting
$\omega _ {\mu 0} = B_ \mu $ (hypercharge-boson).{\footnote{In this context
one has to split off a $g_1$ from the gauge coupling constant $g$.}}
But this will not be performed in this paper in any detail.

\section*{IV. Low Energy Limit and Microscopic Gravity}

In  this section we investigate the gravitational interaction
 between elementary fermionic particles after symmetry-breaking
 under neglection of the very massive $\omega _ {\mu i}$-boson-interaction.
Simultaneously the massless hypercharge
boson is also neglected because it belongs to the range of
 electroweak interactions. Furthermore for simplicity in a first step we
linearize in the following  with respect to
 $\epsilon ^{\mu }{}_ {\nu R}$ and $\epsilon ^\mu {}_ {\nu L}$
(see (3.3a)) under the assumptions that
$|\epsilon ^\mu {}_ {\nu R}| << 1$ and
$|\epsilon ^\mu {}_ {\nu L}| << 1$.
Then the fermionic
2-spinor equations (2.11) using  chapter III
take the form:
$$
i [\sigma ^\mu _R + \epsilon ^\mu {}_ {\nu R} \sigma ^\nu _R]
\partial _ \mu \chi _ {RD} +
 \frac{i}{2} (\partial _\mu  \epsilon ^\mu {}_ {\nu R} ) \sigma  ^\nu  _R
\chi _ {RD}  -
\eqno (4.1)
$$
$$
- m[ 1 + \frac{1}{4} (\epsilon _ {\nu \sigma L} + \epsilon _ {\sigma \nu R} )
\sigma ^\nu _ R \sigma ^\sigma _L ] \varphi _ {LD} = 0
$$
and
$$
i [\sigma ^\mu _L + \epsilon ^\mu {}_ {\nu L} \sigma ^\nu _ L]
\partial _ \mu \varphi _ {LD} + \frac{i}{2} ( \partial _ \mu \epsilon ^\mu
{}_ {\nu L})
\sigma ^\nu _ L \varphi _ {LD} -
$$
$$
- m [1 + \frac{1}{4}( \epsilon _ {\nu \sigma R} + \epsilon _ {\sigma \nu L})
\sigma ^\nu _L \sigma ^\sigma _ R] \chi _{RD} = 0,
\eqno (4.2)
$$
where the right- and left-handed states possess the same
mass $m$ (see (3.5)).

Going over from a spinorial representation of the
Higgs-field
 equations (2.13) and (2.14) to  Lorentz-tensorial
equations we
 multiply these equations after symmetry-breaking
(without
loss of generality) with $\sigma ^\nu  _ {LB} {}^A $ and
$\sigma ^\nu  _ {RB}{}^A$ respectively and obtain using (2.4):
$$
\partial _\alpha \partial ^\alpha \epsilon ^{\mu \nu } _R -
2 \partial _\alpha \partial ^\mu  \epsilon ^{\alpha \nu } _R
- \frac{\mu ^2}{4} (\epsilon _\alpha {}^\alpha {}_ R
+ \epsilon _\alpha {}^\alpha {}_L ) \eta ^{\mu \nu } =
$$
$$
= \frac{8}{v^2} \left\{  T^{\mu \nu } (\varphi _ {LD} ) -
\frac{m}{4} (\varphi _ {LD} ^{\dagger}
\sigma ^\nu _L \sigma ^\mu _ R \chi _ {RD} +
\chi ^{\dagger} _ {RD} \sigma ^\mu _ R \sigma ^\nu _L \varphi _ {LD} )
\right\}
\eqno (4.3)
$$
and
$$
\partial _\alpha \partial ^\alpha \epsilon ^{\mu \nu } _L -
2  \partial_ \alpha  \partial ^\mu  \epsilon ^{\alpha  \nu } _L
- \frac{\mu ^2}{4} (\epsilon _\alpha {}^\alpha {}_R +
\epsilon _\alpha {}^\alpha {}_L ) \eta ^{\mu \nu } =
$$
$$
= \frac{8}{v^2} \left\{  T^{\mu \nu } (\chi _{RD} ) -
\frac{m}{4}
(\varphi _ {LD} ^{\dagger} \sigma ^\mu _L \sigma ^\nu _R \chi _ {RD} +
\chi ^{\dagger} _ {RD} \sigma ^\nu  _R \sigma ^\mu _L
\varphi _ {LD} ) \right\} .
\eqno (4.4)
$$
Herein $T^{\mu \nu } (\chi _ {RD})$ and $T^{\mu \nu }
 (\varphi _ {LD}) $ are the right- and left-handed
energy-momentum tensors (2.19a) and (2.19b) in their
lowest order and $v^{-2}$ plays the role of the
gravitational constant (c.f. (5.34)).

Here a certain cross-interaction between the right- and
left-handed states exists, which is already present in the original equations
(2.13) and (2.14): The energy of the
right-handed states ($T^{\mu \nu } (\chi _ {RD}))$
generates the gravitational fields $\epsilon ^{\mu \nu } _L$ according to
(4.4) which act back on the left-handed
states ($\varphi _{LD})$ in view of (4.2) and vice versa.
However because of the mass-terms in (4.1), (4.2) and (4.3), (4.4) this
cross-interaction picture applies only in
the
massless case $(k = 0)$ exactly. In consequence of this there
exists no neutrino-neutrino interaction  by the microscopic
gravitational $\epsilon ^{\mu \nu }_L $- , $\epsilon ^{\mu \nu } _R $-
fields, if only left-handed neutrinos exist.

\section*{V. Macroscopic Gravity }

We neglect furthermore the gauge-boson interaction. Moreover in the classical
macroscopic limit right- and left-handed
states may be equally represented, i.e.
$$
T^{\mu \nu } (\chi _ {RD} ) =
T^{\mu \nu } (\varphi _ {LD} ) = \frac{1}{2}T^{\mu \nu }
(\psi _D)
\eqno (5.1)
$$
and in consequence of this it is valid according to (4.3)
and (4.4)
$$
\epsilon ^{\mu \nu } _R = \epsilon ^{\mu \nu } _L =
\epsilon ^{\mu \nu }
\eqno (5.2a)
$$
and in view of (3.3a)
$$
h ^{\mu \nu } _R = h ^{\mu \nu } _L = h ^{\mu \nu } .
\eqno (5.2b)
$$
In this and only this case we can define generalized
Dirac-matrices $\, \tilde \gamma ^\mu $ by setting in
view of (2.2):
$$
\, \tilde \gamma ^\mu = h^\mu {}_ \nu
\left(
\begin{array}{ll}
0 & \sigma ^\nu  _ L \\
\sigma ^\nu  _R & 0
\end{array}
\right)
= h ^\mu {}_ \nu \gamma ^\nu .
\eqno (5.3)
$$
For this reason (5.1) is necessary in the classical limit. Then the fermionic
2-spinor equations (4.1) and (4.2)
can be combined to a generalized Dirac-equation for the
4-spinor $\psi _D$ (see (3.4a)):
$$
i \, \tilde \gamma ^\mu \partial _\mu \psi _ D +
\frac{i}{2} (\partial _\mu \, \tilde \gamma ^\mu )
\psi _D -
\frac{m}{4} \, \tilde \gamma ^\mu \, \tilde \gamma _\mu \psi _D = 0.
\eqno (5.4)
$$
Simultaneously by addition of the Higgs-field equations
(4.3)
 and (4.4) we obtain ($\epsilon = \epsilon ^\mu {}_\mu )$:
$$
\partial _ \alpha \partial ^\alpha \epsilon ^{\mu \nu } -
2 \partial_ \alpha \partial ^\mu  \epsilon ^{\alpha \nu }
- \frac{\mu ^2}{2} \epsilon \eta ^{\mu \nu } =
$$
$$
 = \frac{4}{v^2} \left\{  T^{\mu \nu } (\psi _ D ) -
\frac{1}{2} T (\psi _D) \eta ^{\mu \nu } \right\} ,
\eqno (5.5)
$$
where in the lowest order (modulo Dirac equation)
$$
T (\psi _D ) = m (\varphi _ {LD} ^{\dagger} \chi _ {RD} +
\chi _ {RD} ^{\dagger} \varphi _ {LD} )
\eqno (5.5a)
$$
is the trace of Dirac's canonical energy-momentum
tensor $T^{\mu \nu }(\psi _D)$.
Finally the momentum law (2.22) takes in this
classical limit after a longer calculation the very simple form:
$$
\partial _\nu T _ \mu {}^\nu (\psi _D) =
- (\partial _\mu  \epsilon ^\alpha {}_ \beta ) T _\alpha {}^\beta
(\psi _D) +
\frac{1}{2} (\partial _\mu \epsilon ) T(\psi _D) .
\eqno (5.6)
$$

Here the question of a connection of (5.5) and (5.6) to Einstein's
metrical theory of gravity arises. For this we
have to define at first an effective non-Euclidean metric.

\subsection*{a) The Effective Metric}

We define the effective metric by the mass-shell condition following from the
Dirac-equation in the lowest WKB-limit.
For this we insert (5.3) into (5.4) and find (linearized
in $\epsilon ^{\mu \nu })$:
$$
i \gamma ^\mu {\cal{D}}_ \mu \psi _D -
m (1 + \frac{1}{2}\epsilon ) \psi _D = 0
\eqno (5.7)
$$
with the generalized derivative:
$$
{\cal{D}}_ \mu  = \partial _\mu  + \epsilon ^\nu {}_\mu
\partial _\nu +
\frac{1}{2} (\partial _\nu  \epsilon ^\nu {}_\mu ).
\eqno (5.7a)
$$
Iteration of (5.7), elimination of spin-operator
influences because of aspiring to the classical limit
and consequent linearization in $\epsilon ^{\mu \nu } $
give:
$$
{\cal{D}}_ \mu {\cal{D}} ^\mu \psi _D +
m^2 (1 + \epsilon ) \psi _D +
\frac{i}{2} m \gamma ^\mu (\partial _\mu  \epsilon )
 \psi _D = 0
\eqno (5.8)
$$
or after insertion of (5.7a):
$$
(\partial _\mu \partial ^\mu +
2 \epsilon ^{(\nu \mu )} \partial _ \nu  \partial _ \mu
+ \epsilon ^{\nu \mu } {}_ {| \mu } \partial _\nu +
$$
$$
+ \epsilon  ^{\nu \mu } {}_ {|\nu } \partial _ \mu +
\frac{1}{2} \epsilon ^{(\nu \mu )} {}_ {| \nu | \mu } )
\psi _D +
\eqno (5.8a)
$$
$$
+ \frac{m^2}{\hbar^2} (1 + \epsilon ) \psi _D
+ \frac{i}{2} \frac{m}{\hbar} \gamma ^\mu (\partial _ \mu  \epsilon ) \psi _D
= 0,
$$
where we have introduced $\hbar$ explicitely in view of the WKB-method. The
$\gamma ^\mu $-term will vanish in the following because of (5.23), so that
(5.8a) has indeed the structure of a Klein-Gordon equation.

Now we use the WKB-Ansatz
$$
\psi _D = A \epsilon ^{i W/\hbar}
\eqno (5.9)
$$
($A$ 4-spinorial amplitude, $W$  scalar phase-function) and expand $W$ and
$A$ with respect to $\hbar$ as follows:
$$
\begin{array}{lll}
W & = & \sum _ {n = 0}
\stackrel{(n)}{W} \hbar^n , \\
A & = & \sum _ {n = 0} \stackrel{(n)}{A} \hbar^n  .
\end{array}
\eqno (5.9a)
$$
Insertion of (5.9) and (5.9a) into (5.8a) gives in the
lowest order of $\hbar$ (i.e. $\hbar^0$) the Hamilton-Jacobi
equation:
$$
( \eta ^{\mu \nu } +
2 \epsilon ^{(\mu \nu )} ) \stackrel{(0)}{W}_ {| \mu }
\stackrel{(0)}{W} _ {| \nu } -
m^2 (1 + \epsilon ) = 0,
\eqno (5.10)
$$
which is simultaneously the mass-shell condition for
the canonical 4-momentum of the particle:
$$
p _ \mu  = \stackrel{(0)}{W} _ {| \mu } .
\eqno (5.10a)
$$
Consequently, the effective non-Euclidean metric is defined by
$$
g^{\mu \nu } = \eta ^{\mu \nu } ( 1 - \epsilon ) + 2 \epsilon ^{(\mu \nu )}
\eqno (5.11a)
$$
and because of {\footnote {Note that here the indices
are not lowered by $\eta _ {\mu \nu }$ !}} $g_ {\mu \nu } g^{\nu \lambda } =
\delta _\mu  {}^{\lambda } $
$$
g_ {\mu  \nu } = \eta _{\mu \nu } (1 + \epsilon ) - 2
\epsilon _ {(\mu \nu )} ,
\eqno (5.11b)
$$
so that equation (5.10) takes the form of the mass-shell:
$$
g^{\mu \nu } p _\mu  p _\nu  - m^2 = 0.
\eqno (5.12)
$$
We note here that such an effective general metric for
describing gravity is only possible in the classical limit
defined by (5.1) and (5.2).

Finally we derive the equation of motion of the quantum
particle in its classical limit following from (5.12) by
the 4-gradient; one finds in view of (5.10a):
$$
p _ {\alpha | \mu } p _ {\nu } g^{\mu \nu } +
\frac{1}{2} g^{\mu \nu } {}_ {| \alpha } p_ \mu
p _\nu  = 0.
$$
This is exactly the equation of geodesics with the
effective metric $g^{\mu \nu }$ and its Christoffel-symbols
$\left\{ {}^\mu _ {\alpha \beta } \right\} $ as connection coefficients and
can be written in the form\footnote {$||\mu $ denotes
the covariant derivative with respect to the effective
metric and its Christoffel-symbols.}:
$$
p _ {\alpha || \mu } p _ \nu  g^{\mu \nu } = 0.
\eqno (5.13a)
$$
Consequently the effective non-Euclidean space-time is a Riemannian one.
On the other hand we
note that the metric (5.11a) is connected with the
generalized Dirac-matrices (5.3) by the anti-commutator
relation:
$$
2 g^{\mu \nu } {\bf 1} =
\left\{  \, \tilde \gamma ^\mu  , \, \tilde
\gamma ^\nu \right\} (1 - \epsilon ) .
\eqno (5.14)
$$
Thus only if the trace $\epsilon  \equiv 0$ ,
the matrices $\, \tilde \gamma ^\mu $ define a
Clifford-algebra on the effective metric
$g ^{\mu \nu }$. In the next section we shall show
that $\epsilon \equiv 0$ is indeed valid.

\subsection*{b) The Gravitational Field Equations}

In the next step we derive the field equations for the
effective metric (5.11) starting from the Higgs-equations
(5.5). At first we take the trace of (5.5):
$$
\epsilon ^{|\alpha } {}_ {| \alpha } -
2 \mu ^2 \epsilon - 2 \epsilon ^{(\alpha \beta )}
 {}_ {|\alpha | \beta } =
- \frac{4}{v^2} T (\psi _D)
\eqno (5.15)
$$
and eliminate herewith $T(\psi _D)$ in (5.5) giving:
$$
\epsilon ^{\mu \nu | \alpha } {}_ {| \alpha }
- 2 \epsilon ^{\alpha \nu } {}_ {| \alpha } {}^{| \mu } +
(\epsilon ^{(\alpha \beta )} {}_ {| \alpha | \beta } -
\frac{1}{2}\epsilon ^{| \alpha } {}_ {| \alpha } +
\frac{\mu ^2}{2} \epsilon ) \eta ^ {\mu \nu } =
\frac{4}{v^2} T^	{\mu \nu } (\psi _D ) .
\eqno (5.16)
$$
Subsequently we decompose (5.16) into its symmetric
and antisymmetric part resulting in:
$$
\epsilon ^{(\mu \nu ) | \alpha } {}_ {| \alpha } -
\epsilon ^{(\alpha \nu )} {}_ {| \alpha } {}^{| \mu } -
\epsilon ^{[\alpha \nu ] } {}_ {| \alpha } {} ^{|\mu}
- \epsilon ^{(\alpha \mu )} {}_ {| \alpha } {}^{|\nu } -
\epsilon ^{[\alpha \mu ] } {}_ {| \alpha } {}^{| \nu } +
\eqno (5.17)
$$
$$
 + (\epsilon ^{(\alpha \beta )} {}_ {| \alpha |
\beta } -
\frac{1}{2} \epsilon ^{| \alpha } {}_ {| \alpha } +
\frac{\mu ^2}{2} \epsilon ) \eta ^{\mu \nu } =
\frac{4}{v^2} T^{(\mu \nu )} (\psi _D)
$$
and
$$
\epsilon ^{[\mu \nu ]| \alpha } {}_{| \alpha } -
\epsilon ^{(\alpha \nu) } {}_ {| \alpha } {}^{| \mu  } -
\epsilon ^{[ \alpha \nu ]} {}_ {| \alpha } {}^{| \mu } +
\epsilon ^{(\alpha \mu )} {}_ { | \alpha } {}^{| \nu } +
\epsilon ^{[\alpha \mu ]} {}_ {| \alpha } {}^{| \nu } =
\frac{4}{v^2} T^{[\mu \nu ]} (\psi _D ) .
\eqno (5.18)
$$
In the lowest order considered here $(|
\epsilon ^{\mu \nu } | << 1)$ it follows from (5.6):
$$
T^{\mu \nu } ( \psi _D ) _ {| \nu } = 0,
\eqno (5.19)
$$
which decomposes automatically into
$$
T^{(\mu \nu )} (\psi _D ) _ {| \nu } = 0,
\eqno (5.19a)
$$
and
$$
T^{[\mu \nu ]} (\psi _D ) _ {| \nu } = 0
\eqno (5.19b)
$$
in consequence of the fact, that in Dirac's theory also the
 divergence of the symmetrized canonical energy momentum
 tensor vanishes.

Applying at first condition (5.19a) on (5.17) gives
$$
q^{\mu | \nu } {}_ {| \nu } -
\frac{\mu ^2}{2} \epsilon ^{| \mu } \equiv 0
\eqno (5.20)
$$
with the abbreviation:
$$
q^\mu = \epsilon ^{[\alpha \mu ]} {}_ {| \alpha } +
\frac{1}{2} \epsilon ^{| \mu } .
\eqno (5.20a)
$$
Taking in (5.20) the 4-divergence with respect to $x^\mu $, we get
immediately:
$$
\Box ( \epsilon ^{|\mu } {}_ {| \mu } - \mu ^2 \epsilon ) \equiv 0.
\eqno (5.21)
$$
The only solution of (5.21) without any source reads (note that $\epsilon $
vanhishes asymptotically)
$$
\Box \epsilon - \mu ^2 \epsilon \equiv 0,
\eqno (5.22)
$$
which has also the only sourcefree solution
$$
\epsilon \equiv 0.
\eqno (5.23)
$$
The Higgs-mass $(- \mu ^2)$ is connected with the vanishing trace alone, i.e.
the massive Higgs-state is not excited. Herewith equation (5.20) takes the
form:
$$
\Box q^\mu \equiv 0.
\eqno (5.24)
$$
Again the only sourcefree solution is (note that also $q^\mu $ vanishes
asymptotically){\footnote{In view of (5.36) it may be of interest, that also
$q^\mu $ is a 4-gradient: From (5.20) it follows immediately $\Box (q^{\mu
|\nu } - q^{\nu |\mu}) \equiv 0 $ with the only sourcefree solution $q^{\mu
|\nu} - q^{\nu |\mu} \equiv 0$, so that it is valid  $q^ \mu = q^{| \mu }$.
In view of (5.25) it follows $ q = \mbox{const.} = 0$ ( $\mbox{const.} = 0$
because of the boundary condition in the infinity). }}
$$
q^\mu  \equiv 0 ,
\eqno (5.25)
$$
which results in view of (5.20a) and (5.23) in
$$
\epsilon ^{[\alpha \mu ]} {}_ {| \alpha } \equiv 0.
\eqno (5.26)
$$
With (5.23) and (5.26) the first condition (5.20) is fulfilled. The second
condition following by applying (5.19b) on (5.18) reads with the use of
(5.26):{\footnote{The appearance of gauge conditions for the Higgs-fields in
consequence of the conservation laws is a result of the symmetry breaking.}}
$$
(\epsilon ^{(\alpha \mu )} {}_ {| \alpha } {}^{| \nu } -
\epsilon ^{(\alpha \nu )} {}_ {| \alpha } {}^{| \mu }) _ {| \nu }
\equiv 0 .
\eqno (5.27)
$$

Before we investigate  condition (5.27) in more detail we compare our result
with Einstein's theory of gravity. At first we note, that because of (5.23)
the effective metric (5.11) reads finally:
$$
\begin{array}{lll}
g^{\mu \nu } & = & \eta ^{\mu \nu } +
2 \epsilon ^{(\mu \nu )} , \\
g_ {\mu \nu } & = & \eta _ {\mu \nu } - 2 \epsilon _ {(\mu \nu )}
\end{array}
\eqno (5.28)
$$
and (5.14) yields
$$
\left\{  \, \tilde \gamma ^\mu  , \, \tilde \gamma ^\nu  \right\}  =
2 g ^{\mu \nu } {\bf 1} ,
\eqno (5.28a)
$$
defining a Clifford-algebra on the effective metric $g^{\mu \nu }$. Because
of (5.23) and (5.26) the field equations (5.17) and (5.18) take the form:
$$
\epsilon ^{( \mu \nu )| \alpha } {}_ {| \alpha } -
\epsilon ^{(\alpha \nu )} {}_ {| \alpha } {}^{| \mu } -
\epsilon ^{(\alpha \mu )} {}_ {| \alpha } {} ^{| \nu }
+
\epsilon ^{(\alpha \beta) } {}_ {| \alpha | \beta } \, \eta ^{\mu \nu } =
\frac{4}{v^2} T^{(\mu \nu )} (\psi _D )
\eqno (5.29)
$$
and
$$
\epsilon ^{[\mu \nu ] | \alpha } {}_ {| \alpha } +
\epsilon ^{(\alpha \mu )} {}_ {| \alpha } {}^{| \nu } -
\epsilon ^{(\alpha \nu )} {}_ {| \alpha } {}^{| \mu } =
\frac{4}{v^2} T^{[\mu \nu ]} (\psi _D ).
\eqno (5.30)
$$
It is remarkable, that because of (5.23) the Higgs-mass-term proportional to
$\mu ^2$ in (5.17) drops out. The gravitational field is massless, i.e. it
consists of the massless Goldstone-states of the $\, \tilde \sigma
$-Higgs-field alone.

Now we compare (5.29) with Einstein's linearized field equations setting
there
$$
g_ {\mu \nu }  = \eta _ {\mu \nu } + \gamma _ {\mu \nu } \quad (| \gamma _
{\mu \nu }| << 1).
\eqno (5.31)
$$
Comparison with (5.28) gives at first:
$$
\gamma _ {\mu \nu } = - 2 \epsilon _ {(\mu \nu )} .
\eqno (5.31a)
$$
Then the condition (5.23) means
$$
\gamma _ \alpha  {}^{\alpha } \equiv  \gamma  \equiv 0
\Longleftrightarrow \mbox{det} g_ {\mu \nu } \equiv g \equiv -1 .
\eqno (5.32)
$$
Thus we have to take Einstein's field equations in the special
 gauge (5.32), which Einstein has used already in his basic paper
 of 1916 (Einstein-gauge).{\footnote{This gauge has the advantage, that the
non-linear Einstein equations become polynomial, comparable with our
non-linear theory.}} In this gauge Einstein's linearized equations are given
by:
$$
\frac{1}{2} [\gamma _{\mu \nu } {}^{|\alpha } {}_ {| \alpha } -
\gamma _ \nu {}^\alpha {}_ {| \alpha | \mu } - \gamma _ \mu {}^\alpha {}_
{|\alpha | \nu } +
\gamma _ {\alpha \beta } {}^{| \alpha | \beta } \, \eta _ {\mu \nu }] =
- 8 \pi G T_ {(\mu \nu )} .
\eqno (5.33)
$$
Insertion of (5.31a) shows, that equation (5.33) is indentical with (5.29),
if we set:
$$
v^2 = (2 \pi G) ^{-1} .
\eqno (5.34)
$$
The gravitational constant is a consequence of the symmetry breaking! In the
classical limit our theory conincides in its linearized version exactly with
Einstein's linearized theory of gravity.

Let us now consider the integration procedure of the  field equations (5.29)
and (5.30) under the conditions (5.23), (5.26) and  (5.27). For this we
investigate at first the condition (5.27) in more  detail. It means:
$$
\epsilon ^{(\alpha \mu )} {}_ {| \alpha } {}^{| \nu } {}_ {|\nu } =
\epsilon ^{(\alpha \nu )} {}_ {| \alpha | \nu } {}^{| \mu } =
w ^{|\mu }
\eqno (5.35)
$$
$(w = \epsilon ^{(\alpha \nu) } {}_{|\alpha |\nu })$
and therefore it holds:
$$
\Box \epsilon ^{(\alpha \mu )} {}_ {| \alpha } =
w ^{| \mu } .
\eqno (5.35a)
$$
Then it follows immediately:
$$
\Box (\epsilon ^{(\alpha \mu )} {}_ {| \alpha } {}^{|\nu } -
\epsilon ^{(\alpha \nu )} {}_ {| \alpha } {}^{| \mu } )
\equiv 0 ,
\eqno (5.35b)
$$
with the only totally sourcefree solution:
$$
\epsilon ^{(\alpha \mu )} {}_ {| \alpha } {}^{| \nu } -
\epsilon ^{(\alpha \nu )} {}_ {| \alpha } {}^{| \mu }
\equiv 0.
\eqno (5.35c)
$$
Hence $\epsilon ^{(\alpha \mu )} {}_ {| \alpha } $ is a 4-gradient, i.e.
$$
\epsilon ^{(\alpha \mu )} {}_ {| \alpha } = f^{| \mu } .
\eqno (5.36)
$$
Herewith the condition (5.27) is fulfilled.

In consequence of the remaining gauge conditions (5.23), (5.26) and (5.36)
the field-equations (5.29) and (5.30) take the following final form using
(5.34):
$$
\epsilon ^{(\mu \nu ) | \alpha } {}_ {| \alpha } -
2 f^{|\mu |\nu } +
f^{| \alpha }{}_ {|\alpha } \eta ^{\mu \nu } =
8 \pi  G T^{(\mu \nu )} (\psi _D) ,
\eqno (5.37a)
$$
$$
f^{ | \alpha } {}_ {| \alpha } =
4 \pi G T (\psi _D ) ,
\eqno (5.37b)
$$
$$
\epsilon ^{[\mu \nu ]| \alpha } {}_ {| \alpha } =
8 \pi G T^{[ \mu \nu ]} (\psi _D ) .
\eqno (5.37c)
$$
Because of the vanishing divergences of $T^{(\mu \nu )} (\psi _D )$ and
$T^{[\mu \nu ]} (\psi _D)$, see (5.19), all gauge conditions
 are also consequences of the field equations (5.37), which can be integrated
by retarded integrals in a straightforward manner.
 Equation (5.37b) is the Lorentz-invariant Poisson-equation for
 the Lorentz-invariant generalization of the Newtonian gravitational
potential, so that (up to the sign) $f$ has this
 meaning . Of course, the same integration
 procedure is possible for Einstein's field equations (5.33) with the gauge
condition (5.32). Finally we note that for describing  classical
 macroscopic gravity only the field equations (5.37a) and (5.37b) are
necessary. Equation (5.37c) for the
 antisymmetric part of $\epsilon ^{\mu \nu }$ does not possess a classical
analogon in Einstein's theory and $\epsilon ^{[\mu \nu ]}$ plays no role in
the classical limit. However in the complete Dirac-equation (5.7) it appears,
whereas it drops out in the Klein-Gordon equation (5.8a) because of (5.26).
Consequently $\epsilon ^{[\mu \nu ]}$ couples only to spin-properties, which
is confirmed by its source in (5.37c) taking the form in the lowest order:
$$
T^{[\mu \nu ]} (\psi _D) = \frac{1}{2}
\left[ \overline{\psi }_D \gamma ^{[\mu } \sigma ^{\nu ] \lambda } \partial
_\lambda  \psi _D +
(\partial _\lambda  \overline{\psi }_D ) \sigma ^{\lambda [ \mu }
\gamma ^{\nu ]} \psi _D \right] ,
\eqno (5.38)
$$
where $\sigma ^{\mu \nu } = i \gamma ^{[ \mu } \gamma ^{\nu ]} $ is the
spin-operator.

\subsection*{c) The Equations of Motion}

By equation (5.13a) we have already shown that the quantum
 particle in its classical limit moves along geodesics.
Here we prove first that the momentum-law (5.6) is exactly
that of Einstein's theory and secondly that the iterated Dirac-equation
(5.8a) is identical with that in Einstein's theory.

Concerning the momentum law (5.6) we  decompose on the right-hand side the
energy-momentum tensor into its symmetric and antisymmetric part;
because of (5.23) we find:
$$
\partial _\nu T_ \mu  {}^\nu  (\psi _D) =
- (\partial _ \mu  \epsilon _ {(\alpha \beta )})  T^{(\alpha \beta )} (\psi
_D) - (\partial _\mu \epsilon _ {[\alpha \beta ]} ) T^{[\alpha \beta ]} (\psi
_D) .
\eqno (5.39)
$$
Here it is confirmed once more that $\epsilon _ {[\mu \nu ]}$ couples only to
spin-properties, which are also its source (see (5.37c), (5.38)). {\footnote{
The second term on the right hand side of (5.39) and equation (5.37c) are
comparable with the torsion expressions in the Poincar! gauge theory proposed
by Hehl (1973/74). However an interpretation of these supplements as such of
an effective torsion is not possible.}} But because  we have to neglect all
spin-influences within the classical limit, equation (5.39) goes over into:
$$
\partial _\nu T_ \mu {}^\nu (\psi _ D ) = -
(\partial _\mu  \epsilon _ {(\alpha \beta )} )  T^{(\alpha \beta ) } (\psi
_D) .
\eqno (5.39a)
$$
This equation, in which $\epsilon _ {(\alpha \beta )}$ acts back on its
source according to (5.37a), is identical with that in Einstein's
theory for the symmetric energy momentum tensor:
$$
T_ \mu {}^\nu {}_ {||\nu }  = 0
\Rightarrow \partial _\nu  T_ \mu  {}^\nu  =
- \left\{  {}^\alpha _ {\alpha \beta } \right\}  T_\mu {}^\beta   +
 \left\{  {}^\alpha_ {\beta \mu } \right\}  T _\alpha {}^\beta,
\eqno (5.40)
$$
if we take into account:
$$
\begin{array}{l}
 g  =  - 1 \rightarrow  \left\{  {}^\alpha _ {\alpha \beta  } \right\}  = 0,
\\
\left\{  {}^\alpha _ {\beta \mu }\right\}   =
- \epsilon ^{( \alpha }  {}_ {\mu ) | \beta } -
\epsilon _{( \beta } {}^{\alpha )} {}_ {| \mu } +
\epsilon _ {(\beta \mu )} {}^{| \alpha }
\end{array}
\eqno (5.40a)
$$
according to (5.31a) and (5.32).

Secondly, because of the conditions (5.23) and (5.26) the
iterated Dirac-equation (5.8a) takes the form of the
Klein-Gordon equation:
$$
(\partial _\mu  \partial ^\mu
+ 2 \epsilon ^{(\nu \mu )} \partial _ \nu \partial _ \mu  +
2 \epsilon ^{(\nu \mu)} {}_ {| \mu } \partial _\nu
+ \frac{1}{2} \epsilon ^{(\nu \mu )} {}_ {| \nu |\mu }) \psi _D +
m^2 \psi _D = 0.
\eqno (5.41)
$$
Using (5.28) for elimination of $\epsilon ^{(\mu \nu) }$
we get:
$$
(\partial _ \mu \partial _\nu  \psi _D ) g^{\mu \nu }+
g^{\mu \nu } {}_ {| \mu } \partial _\nu  \psi _D
 + \frac{1}{4} g^{\mu \nu } {}_ {| \mu | \nu } \psi _D +
m^2 \psi _D = 0.
\eqno (5.42)
$$
Because of ($R$ Ricci-Scalar)
$$
R = g^{\mu \nu } {}_ {| \mu | \nu } \quad (g = 1)
\eqno (5.42a)
$$
and $\left\{  {}^\alpha _ {\alpha \mu } \right\}  = 0$ (see 5.40a) equation
(5.42) goes over into:
$$
(\psi _ {D| \nu } g^{\mu \nu } ) _ {|| \mu } +
\frac{1}{4} R \psi _D + m^2 \psi _D = 0.
\eqno (5.43)
$$
This is in the vacuum ($R \equiv 0$) the minimally coupled general
 covariant Klein-Gordon  equation. In the framework of
the microphysics the term $ \frac{1}{4} R \psi _D $ plays
no role. Consequently not only on the level of classical
mechanics (equation (5.40)) but also on the quantum
mechanical level (equation (5.43)) our theory is in
accordance with Einstein's gravitational theory within
the linearized versions.

\section*{VI. Conclusions}

We have shown that a ``spin-gauge'' theory of the group $SU(2)
 \times U(1)$ of the 2-spinors representing particle/antiparticle results in
a  gravitational interaction between elementary spin-1/2 particles after
symmetry breaking. The function
valued Pauli-matrices are treated as Higgs-fields and
mediate a gravitational cross-interaction between
 right- and left-handed states. In the classical limit,
where right- and left-handed states are equally
represented, the gravitational interaction can be described
by an effective metric totally in accordance with
Einstein's metrical  theory of  gravity, where we have
restricted ourselves for simplicity to the linearized version of
the theories. The comparison of the non-linear theories is under
investigation.

After symmetry breaking the SU(2)-gauge bosons become very massive of the
order of the Planck-mass and give rise to an additional
"strong gravitational" interaction at very high energies
($ \sim 10^{19} GeV)$ connected with particle-antiparticle transitions; this
is of interest in view of the particle/antiparticle asymmetry in the
Universe. However the U(1)-gauge boson
remains massless, so that it can be identified with that
 of the (weak) hypercharge. Here a unification with the electroweak
 interaction may be possible on the basis of unitary phase
gauge transformations within a high dimensional
(e.g. 4-dimensional) spin-isospin space, which decays
after symmetry breaking into a spin-space and an
isospin-space describing gravitational and electroweak interactions
separately.

We hope that by such a procedure also the chiral asymmetry of
the fermions with regard to the weak interaction, which is
already present in the SU(5)-GUT, can be explained as a
 consequence of the symmetry breaking mentioned above. And
 secondly, the theory, as it stands, describes the
gravitational interaction between fermions only. Within a
complete gravitational theory the interaction with all
bosons must be included, which may also require an
unification with the electroweak interaction. We will
present this in a later paper.
\newpage
\section*{References}

\begin{itemize}
\item[] Bade, W. , Jehle, H. (1953). {\it Review of Modern Physics}, {\bf
25}, 714.
\item[] Barut, A.O. and  McEwan, J. (1984).
 {\it Physics Letters}, {\bf 135}, 172.
\item[] Barut, A.O. and  McEwan, J. (1986).
{\it Letters on Mathematical Physics}, {\bf 11}, 67.
\item[] Dehnen, H. and Hitzer, E. (1994). {\it International Journal of
theoretical Physics}, { \bf 33}, 575.
\item[] Ebner, D. and Dehnen, H. (1993). {\it Physikalische Bl!tter} {\bf
11}, 1013.
\item[] Einstein, A. (1916). {\it Annalen der Physik}, {\bf 49}, 769,
reprinted in Lorentz, H. A., Einstein, A. and Minkowski, H. (1958). {\it Das
Relativit\"atsprinzip}, 6th. ed., Wiss. Buchges. Darmstadt.
\item[] Drechsler, W. (1988). {\it Zeitschrift f\"ur Physik C}, {\bf 41},
197.
\item[] Hehl, F. (1973). {\it General Relativity and Gravitation}, {\bf 5},
333.
\item[] Hehl, F. (1974). {\it General Relativity and Gravitation}, {\bf 4},
491.
\item[] Laporte, O. and Uhlenbeck, B. (1931). {\it Physical Review}, {\bf
37}, 1380.
\item[] Morrison, P. (1958). {\it Am. J. Phys. }, {\bf 26}, 358.
\item[] Nieto, M., Goldman, T. 1991. {\it Phys. Rep.,} {\bf 205}, 221.
\end{itemize}

\newpage

\end{document}